\documentclass[11pt,a4paper,cite]{article}
\usepackage[english]{babel}
\usepackage[T1]{fontenc}
\usepackage[latin1]{inputenc}
\usepackage{epsfig}
\usepackage{graphics}
\usepackage{cite}
\usepackage{amsmath}
\usepackage{amsfonts}
\usepackage{amssymb}
\usepackage{amsthm}
\usepackage
[makeroom]{cancel}
\usepackage{color}
\usepackage{bm}
\usepackage{datetime}
\usepackage{xcolor}

\newcommand{\beq}{\begin{equation}}
\newcommand{\eeq}{\end{equation}}

\newcommand\blfootnote[1]{%
	\begingroup
	\renewcommand\thefootnote{}\footnote{#1}%
	\addtocounter{footnote}{-1}%
	\endgroup
}

\newdateformat{monthyeardate}{%
	\monthname[\THEMONTH], \THEYEAR}

\title{\bf Electromagnetic dual \\ Einstein-Maxwell-scalar models}
\author{Carlos A. R. Herdeiro$^{\ddagger}$, Jo\~{a}o M. S. Oliveira$^{\ddagger,\dagger}$}
\date{%
	{\small $\ddagger$ Departamento de Matem\'atica da Universidade de Aveiro and CIDMA,} \\
	{\small Campus de Santiago, 3810-183 Aveiro, Portugal}\\
	{\small $\dagger$ Centro de Matem\'atica, Universidade do Minho, 4710-057, Braga, Portugal}\\
	\vspace{0.3cm}
	May 2020
}

\begin{document}
\maketitle
\blfootnote{$\ddagger$ herdeiro@ua.pt}
\blfootnote{$\dagger$ jmiguel.oliveira@ua.pt}

\begin{abstract} 
Electromagnetic duality is discussed in the context of Einstein-Maxwell-scalar (EMS) models including axionic-type couplings. This family of models introduces two non-minimal coupling functions $f(\phi)$ and $g(\phi)$, depending on a real scalar field $\phi$.  Interpreting the scalar field as a medium, one naturally defines constitutive relations as in relativistic non-linear electrodynamics. Requiring these constitutive relations to be invariant under the $SO(2)$ electromagnetic duality rotations of Maxwell's theory, defines 1-parameter, closed \textit{duality orbits} in the space of EMS models, connecting different electromagnetic fields in "dual" models with different coupling functions, but leaving both the scalar field and the spacetime geometry invariant. This mapping works as a solution generating technique, extending any given solution of a specific model to a (different) solution for  any of the dual models along the whole duality orbit. We illustrate this technique by considering the duality orbits seeded by specific EMS models wherein solitonic and black hole solutions are known. For dilatonic models, specific rotations are equivalent to $S$-duality.
\end{abstract}

\newpage

\tableofcontents

\section{Introduction}
The parallelism between the laws that rule the electric and magnetic fields $({\bf E},{\bf B})$,  in the absence of sources, is transparent from Maxwell's equations.  In vacuum, these equations are invariant under \textit{electromagnetic duality}:
\begin{equation}
{\bf E}+i{\bf B} \longrightarrow e^{i\beta} ({\bf E}+i{\bf B}) \ ,
\label{md}
\end{equation}
which amounts to an $SO(2)$ rotation by an angle $\beta$.  Two real $\beta$-independent quantities, quadratic in the electromagnetic fields, can be formed, namely:
\begin{equation}
\frac{1}{2}({\bf E}+i{\bf B} )\cdot ({\bf E}+i{\bf B} )^*=\frac{1}{2}{\bf E}^2+{\bf B}^2 \ ,  \qquad \frac{1}{2i}({\bf E}+i{\bf B} )\times ({\bf E}+i{\bf B} )^*={\bf E}\times {\bf B} \ .
\end{equation}
This shows that, despite the change in the fields, electromagnetic duality preserves the electromagnetic energy and momentum densities.

Concrete formulations of electromagnetic duality appeared in the wake  of Maxwell's equations.  In 1893, Heaviside  observed these equations are invariant under the discrete transformation $(\mathbf{E},\mathbf{B}) \rightarrow (-\mathbf{B},\mathbf{E})$~\cite{Heaviside.1893}, which corresponds to \eqref{md} for $\beta=\pi/2$. This invariance was  generalised to the continuous transformation (\ref{md}) by Larmor \cite{Larmor.1928}. It was studied in the context of general relativity by Rainich \cite{Rainich.1925} and revisited by Misner and Wheeler in their attempt to understand classical physics as geometry, wherein the terminology \textit{duality rotation} was introduced~\cite{Misner:1957mt}. In its relativistic formulation, \eqref{md} can be expressed as
\begin{equation}
{\bf F} \longrightarrow \cos\beta  {\bf F} + \sin\beta  \tilde{\bf F} \ ,
\label{mdr}
\end{equation}
where ${\bf F}$ denotes the Maxwell 2-form and $\tilde{\bf F}$ denotes its Hodge dual. This formulation makes clear that duality rotations remain a symmetry of Maxwell's equations in curved spacetime: the covariant theory remains \textit{self-dual}.

Electromagnetic duality rotations are not ordinary rotations in 3-space. They define an equivalence class of electromagnetic fields; that is, there are different  $({\bf E},{\bf B})$ solutions to Maxwell's equations which have the same energy and momentum density. 
However, they are only an invariance of Maxwell's equations in vacuum. For instance, applying \eqref{mdr} with $\beta=\pi/2$ to the electric field of a static, point electric charge $Q$, in standard spherical coordinates in flat spacetime $(t,r,\theta,\varphi)$, leads to
\begin{equation}
{\bf F}=\frac{Q}{r^2} dt\wedge dr \longrightarrow \tilde{\bf  F}=-Q\sin\theta d\theta \wedge d\varphi \ ,
\label{mm}
\end{equation}
which is the field of a static, magnetic monopole, with magnetic charge $Q$. Thus, preserving the duality in the presence of electric charges requires magnetic monopoles. From a different reasoning, Dirac noted that the existence of magnetic monopoles could explain electric charge quantisation~\cite{Dirac.1931}. Up to now, however, magnetic monopoles have no observational support, and thus electromagnetic duality is an unbroken symmetry in vacuum  only. This example illustrates how the $\beta=\pi/2$ rotation, corresponding to the discrete symmetry observed by Heaviside, exchanges electric and magnetic fields.

It is interesting to consider how duality rotations are affected if one generalises Maxwell's theory, modifying its equations of motion. Gibbons and Rasheed considered the case of relativistic non-linear electrodynamics~\cite{GibRas1996}. They obtained the conditions under which a theory of non-linear electrodynamics, possibly coupled to gravity, has invariant equations of motion under duality rotations, and observed this is the case for  Born-Infeld theory~\cite{Born:1934gh}.  This is a rather exceptional theory, see $e.g.$~\cite{Gibbons:2000xe}, which naturally appears as the effective field theory describing open string excitations in  string theory~\cite{Fradkin:1985qd}.  In this context, a low energy effective field theory is an Einstein-Maxwell-dilaton-axion model, where the dilaton is a scalar field and the axion a pseudo-scalar field. In~\cite{GibRas1996} it was shown this model is still self-dual under electromagnetic duality rotations as long as the axion and dilaton mix in an appropriate way under this transformation. Thus, electromagnetic duality maps solutions of the Einstein-Maxwell-dilaton-axion equations to different solutions of the same model - see also~\cite{GibRas1995,GailZum1981,GailZum1997}.

There is, however, a broader notion of duality. Instead of considering self-dual models, which are left invariant (at least at the level of the equations of motion), by some transformation, we can consider \textit{dual theories}: two different models related by a non-trivial duality map. Considering dual theories has been particularly rewarding when the mapping is a strong-weak coupling one. This allows relating a model in the weak coupling regime, wherein perturbative computations are possible, to a technically more challenging strongly coupled model, potentially extracting non-trivial information from the latter. Famous examples include the Sine-Gordon -- Thirring duality~\cite{Coleman:1974bu}, S-duality in string theory~\cite{Sen:1994fa} and, of course, AdS-CFT~\cite{Maldacena:1997re}. The duality map, moreover, can be used at the level of specific solutions, as a means to obtain a solution of one of the models from a known solution of the dual model. In fact, it is often a non-trivial and useful solution generating technique.

In this paper we shall consider a family of models for which electromagnetic duality provides a simple realisation of "dual theories". Then, we shall use this mapping as a solution generating technique. The  family is the Einstein-Maxwell-scalar (EMS) class of models, whose action reads
\beq
\mathcal{S}=\frac{1}{4\pi}\int d^4x \sqrt{-g} \left(\frac{R}{4}-\frac{f(\phi)}{4}F_{\mu\nu}F^{\mu\nu}+\frac{g(\phi)}{4}F_{\mu\nu}\tilde{F}^{\mu\nu}-\frac{1}{2}\partial_\mu\phi \partial^\mu \phi\right) \ ,
\label{action}
\eeq
where $F_{\mu\nu},\tilde{F}_{\mu\nu}$ denote the components of ${\bf F},\tilde{\bf F}$, $\phi$ is a scalar field, $R$ is the Ricci scalar of the metric ${\bf g}$ and $f(\phi)$ and $g(\phi)$ denote unspecified coupling functions. In the following we shall denote a solution of \eqref{action}, for a specific choice of $f(\phi),g(\phi)$ as
\beq
[{\bf g}, {\bf A}, \phi; f(\phi),g(\phi)] \ ,
\label{sol}
\eeq
where ${\bf F}=d{\bf A}$.

Many interesting special cases of the family of models \eqref{action} have been considered in the literature. For instance, in string theory, supergravity and Kaluza-Klein theory one often finds dilatonic couplings, $f(\phi)\sim e^{a\phi}$; the axion coupling, $g(\phi)\sim \phi $, is motivated by the Peccei-Quinn proposal to solve the strong CP problem~\cite{Peccei:1977hh}; recent work on spontaneous scalarisation motivates couplings obeying $df/d\phi(\phi=0)=0$, starting with~\cite{CHetAl.SclChr.2018}.

We shall establish an electromagnetic duality transformation defined by an angle $\beta$, $\mathcal{D}_\beta$, that maps any solution \eqref{sol} of a certain EMS model \eqref{action} to a different solution of a different (dual) model, within the same family,
\beq
[{\bf g}, {\bf A}, \phi; f(\phi),g(\phi)]  \stackrel{\mathcal{D}_\beta}{\longrightarrow} [{\bf g}, {\bf A}', \phi; f_\beta(\phi),g_\beta(\phi)] \ .
\label{orbit}
\eeq 
The rotation angle $\beta$ parameterises orbits in the space of EMS models, that we shall call \textit{duality orbits}. This space is spanned by the functions $f,g$. The orbits are closed and relate dual models. On the one hand, the electromagnetic variables and the couplings $f,g$ are affected by the mapping, transforming from the original $ {\bf A}$ and $f(\phi),g(\phi)$ to a new $ {\bf A}'$ and $f_\beta(\phi),g_\beta(\phi)$, all of which  depend on $\beta$. On the other hand, the metric and the scalar field shall remain invariant along the whole duality orbit. In particular, we shall consider the duality orbits passing through some EMS models recently studied, wherein  black hole solutions \cite{CHetAl.SclChr.2018,PFetAl.SclChr.2019,DAetAl.EMSBH.2019,PFetAl.AxChr.2019,JBSetAl.EMSBH.2020} and solitonic solutions \cite{CHJOER2020} (see also \cite{CHJO2019.1,CHJO2019.2}) have been found

This paper is organised as follows. In section $3$, we present our formulation of the electromagnetic duality for the EMS model, establishing the map between different solutions in models with different coupling functions, preserving the metric and the scalar field. This defines the duality orbits. In section $4$ we present some examples wherein this duality is applied, using explicitly known solutions of illustrative EMS models, to the obtain duality orbits. Conclusions and remarks are presented in section $5$.

\section{Electromagnetic duality in the EMS model}

\subsection{Fields and equations of motion}

Consider the EMS family of models described by the action~\eqref{action}. We shall be interested in stationary asymptotically flat spacetimes, with associated asymptotically timelike Killing vector field $k^\mu$. The scalar field can be regarded as endowing spacetime with a medium, making the electric permittivity and the magnetic permeability spacetime dependent. Then, one uses the standard formalism for electrodynamics in a medium, definining the electric intensity, magnetic induction, electric induction and magnetic intensity 4-covectors as, respectively:\footnote{The fields $D$ and $H$ here correspond to the fields $E'$ and $B'$ in \cite{CHJO2019.2}.}
\begin{align}
E_\mu &= k^\nu F_{\mu\nu}\label{Edef} \ ,\\
B_\mu &= \frac{1}{2}\varepsilon_{\mu\alpha\beta\nu} F^{\alpha\beta} k ^\nu\label{Bdef} = k^\nu \tilde{F}_{\mu\nu} \ ,\\
D_\mu &= k^\nu G_{\mu\nu}(\phi) \label{Drel}\ ,\\
H_\mu &= \frac{1}{2}\varepsilon_{\mu\alpha\beta\nu} G^{\alpha\beta}(\phi) k ^\nu =  k^\nu \tilde{G}_{\mu\nu} \label{Hrel} \ ,\\
\end{align}
where 
\beq
G^{\mu\nu}(\phi) \equiv  -\frac{\partial \mathcal{L}}{\partial F_{\mu\nu}} = f(\phi)F^{\mu\nu} - g(\phi)\tilde{F}^{\mu\nu} \ ,
\eeq
 and $\mathcal{L}$ is the Lagrangian density. The matter equations of motion obtained from \eqref{action} read:
\begin{align}
&\nabla_{[\mu}E_{\nu]}=0, \label{Eeq} \\
&\nabla_{[\mu}H_{\nu]}=0, \\
&\nabla_{\mu}\bigg(\frac{D^\mu}{V}\bigg) = 0, \\
&\nabla_{\mu}\bigg(\frac{B^\mu}{V}\bigg) = 0, \label{Beq} \\
&\Box^2\phi = \frac{1}{4}\frac{d f(\phi)}{d\phi}F_{\mu\nu}F^{\mu\nu}-\frac{1}{4}\frac{dg(\phi)}{d\phi}\tilde{F}_{\mu\nu}F^{\mu\nu}, \label{Phieq}
\end{align}
where $V=-k^\mu k_\mu$ is the norm of the Killing vector field.

\subsection{Constitutive relations}
For electrodynamics in a medium, the constitutive relations specify how the electric and magnetic inductions relate to the electric and magnetic intensities. From  relations \eqref{Drel} and \eqref{Hrel}, the fields $E_{\mu}$ and $H_{\mu}$ are related to $D_{\mu}$ and $B_{\mu}$ fields through the following constitutive relations:
\beq\label{Meq}
\begin{pmatrix}
	E \\
	H
\end{pmatrix} = \frac{1}{f}
\begin{pmatrix}
	1 & g \\
	g & f^2+g^2
\end{pmatrix}
\begin{pmatrix}
	D \\ B
\end{pmatrix}  = M \begin{pmatrix}
	D \\ B
\end{pmatrix}\ .
\eeq
$M$ shall be called the \textit{constitutive matrix}. For $f=1$ and $g=0$, $M$ becomes the identity matrix, and we recover standard vacuum electrodynamics, with $E=D$ and $H=B$ (recall we use units with $c=1$). In general, however, $E,H$ depend on both $D,B$. This is typically the case in non-linear materials and non-linear optics. Thus, one may envisage the non-minimally coupled scalar field as endowing spacetime with a non-linear material environment.

\subsection{Duality map}
We are interested in finding a duality transformation $\mathcal{D}_\beta$ that keeps equations \eqref{Eeq}-\eqref{Phieq} invariant in an appropriate sense. We consider duality $SO(2)$ rotations, by an angle $\beta$, acting on both the intensities and the inductions in the same way, namely~\cite{GibRas1996}:
\beq\label{Duality1}
\begin{pmatrix}
	E \\
	H
\end{pmatrix} \stackrel{\mathcal{D}_\beta}{\longrightarrow}
\begin{pmatrix}
E' \\
H'
\end{pmatrix} =
 S
\begin{pmatrix}
	E \\
	H
\end{pmatrix} \ ,
\eeq
\beq
\begin{pmatrix}
	D \\
	B
\end{pmatrix} \stackrel{\mathcal{D}_\beta}{\longrightarrow}
\begin{pmatrix}
	D' \\
	B'
\end{pmatrix} =
 S
\begin{pmatrix}
	D \\
	B
\end{pmatrix} \ ,
\eeq
where
\beq
S
= 
\begin{pmatrix}
	\cos\beta & \sin\beta \\
	-\sin\beta & \cos\beta
\end{pmatrix} \ ,
\eeq
or, equivalently,
\beq\label{Fdual}
F_{\mu\nu}\stackrel{\mathcal{D}_\beta}{\longrightarrow} F'_{\mu\nu} = \cos\beta \, F_{\mu\nu} + \sin\beta \, \tilde{G}_{\mu\nu} \ ,
\eeq
\beq
G_{\mu\nu}\stackrel{\mathcal{D}_\beta}{\longrightarrow} G'_{\mu\nu} = \cos\beta \, G_{\mu\nu} + \sin\beta \, \tilde{F}_{\mu\nu} \ .
\eeq
Comparing \eqref{Fdual} with \eqref{mdr} one observes this is the standard duality rotation of Maxwell's theory. 
From \eqref{Meq}, it follows that the constitutive matrix becomes
\beq \label{ConstTransf}
M \stackrel{\mathcal{D}_\beta}{\longrightarrow} M' = S M S^{-1} \ ,
\eeq
which reads, explicitly
\begin{align}
M' 
=\frac{1}{f}
\begin{pmatrix}
	f^2\sin^2\beta +(g\sin\beta+\cos\beta)^2 & g\cos(2\beta)+(f^2+g^2-1)\sin(2\beta)/2 \\
	g\cos(2\beta)+(f^2+g^2-1)\sin(2\beta)/2  & f^2\cos^2\beta +(g\cos\beta-\sin\beta)^2
\end{pmatrix} \ .
\label{cm1}
\end{align}
Thus, the duality rotation with an arbitrary angle $\beta$ yields this new  constitutive matrix. The duality orbit of models is defined as the continuous sequence of EMS models~\eqref{action} where  the coupling functions are
\beq
(f(\phi),g(\phi))\stackrel{\mathcal{D}_\beta}{\longrightarrow} (f_\beta(\phi),g_\beta(\phi)) \ ,
\label{dcf}
\eeq
such that
\begin{align}
M' &=\frac{1}{f_\beta}\begin{pmatrix}
	1 & g_\beta \\
g_\beta & f_\beta^2+g_\beta^2
\end{pmatrix} \ .
\label{cm2}
\end{align}
That is, the constitutive relations have the same functional form in terms of the coupling functions, along the whole duality orbit. Comparing \eqref{cm1} with \eqref{cm2} yields
\beq
f_\beta = \frac{f}{f^2\sin^2\beta +(g\sin\beta+\cos\beta)^2} \ ,
\label{fbeta}
\eeq
\beq
g_\beta =\frac{1}{2}\frac{2g\cos(2\beta)+(f^2+g^2-1)\sin(2\beta)}{f^2\sin^2\beta +(g\sin\beta+\cos\beta)^2} \ .
\label{gbeta}
\eeq
The orbit of dual theories is therefore the 1-parameter family of actions
\beq
\mathcal{S}_\beta=\frac{1}{4\pi}\int d^4x \sqrt{-g} \left(\frac{R}{4}-\frac{1}{4}f_\beta(\phi)F'_{\mu\nu}F'^{\mu\nu}+\frac{1}{4}g_{\beta}(\phi)F'_{\mu\nu}\tilde{F}'^{\mu\nu}-\frac{1}{2}\partial_\mu\phi \partial^\mu \phi\right) \ ,
\label{action2}
\eeq
where $\mathcal{S}_0$ equals the original action~\eqref{action} and $({\bf g},{\bf A}',\phi)$, where $ {\bf F}'=d{\bf A}'$, are taken as the independent fields in a variational principle. The tensor $G'_{\mu\nu}$ is found, as before, by the variation of the Lagrangian density in \eqref{action2}, $ \mathcal{L}_\beta$, with respect to $ {\bf F}'$:
\beq\label{Geq}
G'^{\mu\nu} (\phi) \equiv  -\frac{\partial \mathcal{L}_\beta}{\partial F'_{\mu\nu}} = f_\beta(\phi) F'^{\mu\nu}-\frac{1}{4}g_\beta(\phi)\tilde{F}'^{\mu\nu}\ .
\eeq

From the discussion above, it follows that if 
\beq
[{\bf g}, {\bf A}, \phi; f(\phi),g(\phi)] \ , 
\label{sol1}
\eeq 
 is a solution of \eqref{action}, then 
\beq
[{\bf g}, {\bf A}', \phi; f_\beta(\phi),g_\beta(\phi)]
\eeq
 is a solution of the Maxwell equations obtained from \eqref{action2}. It remains to check the scalar and Einstein equations are also obeyed for the model \eqref{action2}.

The scalar equation of motion derived from \eqref{action2} is
\begin{equation}
\Box^2\phi = \frac{1}{4}\frac{df_\beta}{d\phi}F'_{\mu\nu}F'^{\mu\nu}  -\frac{1}{4} \frac{dg_\beta}{d\phi}\tilde{F}'_{\mu\nu}F'^{\mu\nu} \ .
\label{newse}
\end{equation}
Using the identities $F'_{\mu\nu}F'^{\mu\nu} = 2({\mathbf B'}^2-{\mathbf E'}^2)/V$ and $\tilde{F}'_{\mu\nu}F'^{\mu\nu} = -4 {\mathbf E'}\cdot {\mathbf B'}/V$ and then reverting back to the original fields, it follows \eqref{newse} reduces to the original equation of motion \eqref{Phieq} for the scalar field, which is obeyed, since \eqref{sol1} is a solution of \eqref{action} by assumption.

It is also easy to check that the Einstein equations of \eqref{action} and \eqref{action2} are the same. The energy-momentum  tensor of the model \eqref{action2}, $T'_{\mu\nu}$ is obtained from the action by differentiating with respect to the metric, which is unchanged by the duality rotation. Then, the functional form of the energy-momentum tensor is the same as that of the model~\eqref{action}, $T_{\mu\nu}$, and they are mapped simply replacing $({\bf A},f,g)\rightarrow ({\bf A}',f_\beta,g_\beta)$. One  can then show  that 
\beq
T'_{\mu\nu} = T_{\mu\nu} \ ,
\eeq
 by a straightforward application of the transformations. We have thus established the duality orbit of solutions \eqref{orbit}, under \eqref{Fdual} and \eqref{dcf}, the latter explicitly given by \eqref{fbeta}-\eqref{gbeta}.

\begin{figure}[ht!]
\begin{center}
\includegraphics[height=.35\textheight, angle =0]{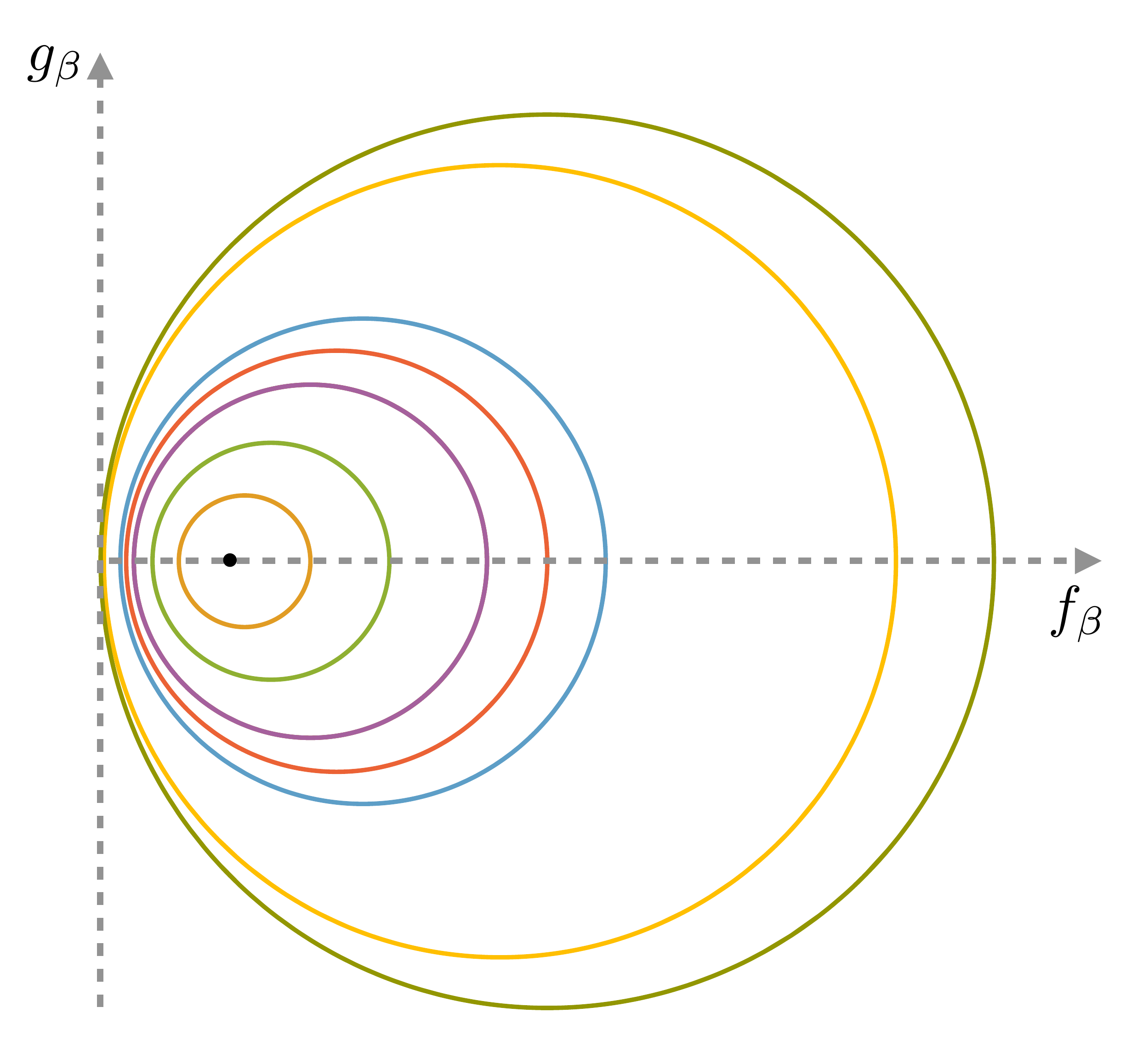}
\end{center}
  \vspace{-0.5cm}
\caption{Duality orbits in the space of EMS models, for different values of the fiducial EMS model $(f,g)$. The self-dual model, Maxwell's theory, is the black dot at $(1,0)$.}
\label{fig1}
\end{figure}

A represention of the duality orbits is obtained as follows. Consider a two dimensional space parameterised by $(x,y)=(f_\beta,g_\beta)$ as an illustration of the space of EMS models. It is simple to check that the duality orbits defined by \eqref{fbeta}-\eqref{gbeta} obey:
\beq
(f_\beta-A)^2+g_\beta^2= A^2-1 \ , \qquad {\rm where} \quad A\equiv \frac{1+f^2+g
^2}{2f} \ .
\eeq
Thus, they are circles, passing through the fiducial EMS model $(f_0,g_0)=(f,g)$. The radius of the circles vanishes at the self-dual model $(f_0,g_0)=(1,0)$, that is, Maxwell's theory. This is illustrated in Fig.~\ref{fig1}.

\section{Examples of duality orbits}

\subsection{Closed form solution for a scalarised electric charge in flat spacetime} \label{toysol}
Our first example is a scalarised electric charge solution found in \cite{CHetAl.SclChr.2018}, for the model \eqref{action} in flat spacetime and with coupling functions
\beq 
f(\phi)=\frac{1}{1-\phi^2} \ , \qquad  g(\phi)=0 \ .
\label{ex1}
\eeq
The scalar field and electric potential are radial functions
\beq
\phi(r) = \zeta \sin\bigg(\frac{Q}{r}\bigg) \ , \qquad   V(r) =\frac{Q}{r}+\zeta^2\bigg[\frac{1}{4}\sin\bigg(\frac{2Q}{r}\bigg)-\frac{Q}{2r}\bigg] \ ,
\label{ex1s}
\eeq
from which the electric intensity and induction have only the radial component:
\beq
E_r = \frac{Q}{r^2}\bigg[1-\zeta^2\sin^2\bigg(\frac{Q}{r}\bigg)\bigg] \ , \qquad 
D_r = f(\phi)E_r = \frac{Q}{r^2} \ ,
\label{ex1e}
\eeq
whereas the magnetic induction and intensity vanish
\beq
\mathbf{B} = 0 = \mathbf{H} \ .
\label{ex1b}
\eeq
Observe that whereas the electric intensity is sensitive to the scalar field, the electric induction has the standard Coulombian form, and it is the same as when $\zeta=0$.

The duality orbit that goes through the model \eqref{ex1} has:
\beq
f_\beta=\frac{1-\phi^2}{1-2\cos^2\beta\phi^2+\cos^2\beta\phi^4} \ , \qquad g_\beta=\frac{\phi^2(2-\phi^2)\sin\beta\cos\beta}{1-2\cos^2\beta\phi^2+\cos^2\beta\phi^4} \ .
\eeq
Along this sequence of dual models,  the seed \eqref{ex1s}-\eqref{ex1b} is mapped, generically, to dyonic solutions. For an arbitrary $\beta$,  the fields along this orbit are:
\beq
E'_r =  \frac{Q}{r^2}\bigg[1-\zeta^2\sin^2\bigg(\frac{Q}{r}\bigg)\bigg]\cos\beta \ , \qquad
D'_r = \frac{Q}{r^2}\cos\beta \ ,
\eeq
\beq
B'_r = -\frac{Q}{r^2}\sin\beta \ , \qquad
H'_r = -\frac{Q}{r^2}\bigg[1-\zeta^2\sin^2\bigg(\frac{Q}{r}\bigg)\bigg]\sin\beta \ .
\eeq
Again, one observes the Coulombic form of the electric and magnetic induction fields, with electric and magnetic charges, respectively, $Q_\beta\equiv Q\cos\beta$ and $P_\beta\equiv Q\sin\beta$, such that
\beq
Q_\beta^2+P_\beta^2\equiv Q^2 = {\rm constant} \ ,
\eeq
along the whole duality orbit. 

Within this orbit there is, however, a pure magnetic solution at $\beta=\pi/2$, wherein the coupling functions are 
\beq
f_{\pi/2}(\phi) =\frac{1}{f(\phi)} = 1-\phi^2 \ ,   \qquad  g_{\pi/2}(\phi)=0 \ ,\label{couppi2}
\eeq
the electric intensity and induction vanish
\beq
\mathbf{E}' = 0 = \mathbf{D}' \ ,
\eeq
and the magnetic induction and intensity are only radial functions:
\beq
B'_r = -\frac{Q}{r^2} \ , \qquad
H'_r = f_{\pi/2}(\phi) B_r =  -\frac{Q}{r^2}\bigg[1-\zeta^2\sin^2\bigg(\frac{Q}{r}\bigg)\bigg] \ .
\eeq
We thus found a pure magnetic solution for the model with couplings \eqref{couppi2}. The original electric charge $Q$ becomes the magnetic charge just as in the Maxwell theory example \eqref{mm}. For $\beta=\pi$ we would get the original purely electric solution but with opposite charge sign while for $\beta=3\pi/2$ we get the pure magnetic solution once again with opposite charge sign. For any $\beta$ value between these, we get a dyon whose magnetic and electric charges relative contributions depend on how close $\beta$ is to the values mentioned above. There is a full orbit of solutions that can be obtained from the original solution.

Let us close this example with two observations. First, this formalism unveils the fact that although the original solution has a non-Coulombian electric intensity, the electric induction is Coulombian. The same holds along the whole duality orbit. Second, at $\beta=\pi/2$ the $f(\phi)$ coupling function is mapped into its inverse,  whereas $g(\phi)$ remains zero. This is a generic feature starting with arbitrary $f(\phi)$ and vanishing $g(\phi)$, as can be appreciated from \eqref{fbeta}-\eqref{gbeta}:
\beq
f_\beta  \ \stackrel{\beta=\pi/2}{=} \   \frac{f}{f^2 +g^2}  \   \stackrel{g=0}{=}  \ \frac{1}{f} \ ,
\label{fbetapi2}
\eeq
\beq
g_\beta \  \stackrel{\beta=\pi/2}{=}\ -\frac{g}{f^2 +g^2}  \   \stackrel{g=0}{=}  \ 0 \ .
\label{gbetapi2}
\eeq
 Since $f(\phi)$ defines the coupling strength of the Maxwell field, this particular value of the map is an example of a strong $\leftrightarrow$ weak coupling duality, with an electric $\leftrightarrow$ magnetic mapping, reminscent of the Montonen-Olive duality~\cite{Montonen:1977sn}.

\subsection{Closed form Maxwell-scalar solitons in flat spacetime}
Our second example uses a seed configuration found in \cite{CHJOER2020}. It describes a purely electric, static, spherically symmetric soliton solution of \eqref{action} in flat spacetime, with
\beq
f(\phi)= \frac{1}{(1-\alpha\phi)^{4}} \ , \qquad g(\phi)=0 \ .
\label{ex2}
\eeq
The scalar field reads
\beq
\phi=\frac{Q}{\alpha Q + r} \ ,
\label{ex2s}
\eeq
the electric intensity and induction have again only a radial component
\beq\label{CharQ}
E_r = \frac{Qr^2}{(r+\alpha Q)^4} \ , \qquad  D_r = f(\phi)E_r = \frac{Q}{r^2} \ ,
\eeq
whereas the magnetic induction and intensity again vanish 
\beq
\mathbf{B} = 0 = \mathbf{H} \ .
\label{ex2b}
\eeq
The $f(\phi)$ coupling \eqref{ex2} diverges at the origin $r=0$; but all physical quantities are regular, such as the energy density and the electric intensity. Indeed, this solution was interpreted in~ \cite{CHJOER2020} as a de-singularisation of the Coulomb solution of Maxwell's theory. Nonetheless, the electric induction $\mathbf{D}$ is again Coulombian and diverges at the origin.

The duality orbit that goes through the model \eqref{ex2},  has:
\beq
f_\beta=\frac{(1-\alpha\phi)^4}{\sin^2\beta+\cos^2\beta(1-\alpha\phi)^8} \ , \qquad g_\beta=\frac{[1-(1-\alpha\phi)^8]\sin\beta\cos\beta}{\sin^2\beta+\cos^2\beta(1-\alpha\phi)^8} \ .
\eeq
Once more, the duality map will generate dyonic solutions from the seed \eqref{ex2s}-\eqref{ex2b}. The fields for this orbit are:
\beq
E'_r =  \frac{Qr^2}{(r+\alpha Q)^4}\cos\beta \ , \qquad
D'_r = \frac{Q}{r^2}\cos\beta \ ,
\eeq
\beq
B'_r = -\frac{Q}{r^2}\sin\beta \ , \qquad
H'_r = -\frac{Qr^2}{(r+\alpha Q)^4}\sin\beta \ .
\eeq
A pure magnetic solution is obtained $\beta=\pi/2$. The dual model at this $\beta$ value has couplings 
\beq
f_{\pi/2}(\phi)=\frac{1}{f(\phi)} = (1-\alpha\phi)^4 \ , \qquad g_{\pi/2}(\phi)=0 \ , 
\label{ex2d}
\eeq
and the dual configuration has vanishing electric intensity and induction
\beq
\mathbf{E}' = 0 = \mathbf{D}' \ ,
\eeq
and a spherical magnetic induction and intensity
\beq
B'_r = -\frac{Q}{r^2} \ , \qquad
H'_r = f_{\pi/2}(\phi) B'_r = -\frac{Qr^2}{(r+\alpha Q)^4} \ .
\eeq
The magnetic induction of the dual solution $B'_r$ is Coulombic and diverges at $r=0$ while $\mathbf{H}'$ is regular. All physical quantities are regular, including the energy density, making this a regular magnetic soliton.

A remark is in order, concerning the no-go theorems in \cite{CHJO2019.2} for strictly stationary solitons. It was shown in \cite{CHJO2019.2} that, as long as the couplings $f(\phi)$ and $g(\phi)$ do not diverge, there are no solitonic solutions. However, a regular solitonic solution with a diverging $f(\phi)$ can be duality rotated into a solution of a model wherein the new coupling does not diverge. In this example, the new non-minimal coupling \eqref{ex2d} actually vanishes instead of diverging, when the original coupling $f(\phi)$ diverges. This solution circumvents the soliton no-go theorem by being in the duality orbit of a solution with divergent coupling.

\subsection{Closed form dilatonic solution in flat spacetime}
As yet another example, consider the spherically symmetric solution discussed in \cite{CHJOER2020} for a dilatonic coupling, $f = e^{-\alpha\phi}$, in flat spacetime. The scalar field and electric potential read
\beq
\phi = -\frac{2}{\alpha}\ln\bigg(1+\frac{\alpha Q}{2r}\bigg) \ , \qquad V(r) = -\frac{2Q}{\alpha Q + 2r} \ .
\eeq
whereas the electric intensity and induction fields are
\beq
E_r = \frac{4Q}{(\alpha Q + 2r)^2} \ , \qquad D_r = f(\phi)E_r = \frac{Q}{r^2} \ .
\eeq
The magnetic induction and intensity are trivial
\beq
\mathbf{B} = 0 = \mathbf{H} \ .
\eeq

In this case, the duality orbit that goes through this model,  has:
\beq
f_\beta=\frac{1}{e^{\alpha\phi}\cos^2\beta+e^{-\alpha\phi}\sin^2\beta} \ , \qquad g_\beta=-\frac{\sin{2\beta}\sinh{\alpha\phi}}{e^{\alpha\phi}\cos^2\beta+e^{-\alpha\phi}\sin^2\beta} \ .
\label{dod}
\eeq
The fields obtained from the seed solution are, along the duality orbit,
\beq
E'_r =  \frac{4Q}{(\alpha Q + 2r)^2}\cos\beta \ , \qquad
D'_r =  \frac{Q}{r^2}\cos\beta \ ,
\eeq
\beq
B'_r = -\frac{Q}{r^2}\sin\beta \ , \qquad
H'_r = -\frac{4Q}{(\alpha Q + 2r)^2}\sin\beta \ .
\eeq
The reasoning is the same and we can see there is, once again, a magnetic solution for $\beta=\pi/2$ with trivial electric intensity and induction.

%

\subsection{The GMGHS black hole}
\label{gmghs}
We now consider a curved spacetime generalisation of the example in the last subsection. This is the well known dilatonic electrically charged, spherically symmetric black hole (in four spacetime dimensions), obtained in the model \eqref{action} with
\beq 
f(\phi)=e^{-2\phi} \ , \qquad g(\phi)=0 \ ,
\eeq
It was first discussed by Gibbons and Maeda in \cite{GibMae.1988} and later by Garfinkle, Horowitz and Strominger~\cite{GarHoroStrom.1991}. We shall call it the GMGHS black hole. The metric reads
\beq\label{lelGM}
ds^2 =-\bigg(1-\frac{2M}{r}\bigg)dt^2+\bigg(1-\frac{2M}{r}\bigg)^{-1}dr^2 + r^2\bigg(1-\frac{r_-}{r}\bigg)(d\theta^2+\sin^2\theta d\varphi^2) \ ,
\eeq
where $M$ is the black hole mass, $Q$ is the electric charge, $ r_- = e^{2\phi_\infty}Q^2/M$ and $\phi_\infty$ is the asymptotic value of the scalar field. The scalar field and the gauge potential read
\beq
e^{2\phi} = e^{2\phi_\infty}\bigg(1-\frac{r_-}{r}\bigg)\ , \qquad {\bf A}=-\frac{Q}{r}e^{2\phi_\infty} {dt} \ ,
\label{gmghss}
\eeq
whereas the electric intensity and induction are
\beq
E_r = \frac{Q}{r^2}e^{2\phi_\infty} \ , \qquad
D_r = f(\phi)E_r = \frac{Q}{r(r-r_-)} \ ,
\eeq
and the magnetic induction and intensity vanish:
\beq
\mathbf{B} = 0 = \mathbf{H} \ .
\eeq
We remark that ${\bf D}$ does not have a Coulombic form, unlike the above cases. This is because the radial coordinate in \eqref{lelGM} is not the areal radius. Using the areal radius $r^*= r\sqrt{1-r_-/r}$, the Coulombic form $D_r= Q/r^{*2}$ is recovered.

The duality orbit that goes through this model,  has the form \eqref{dod} with $\alpha=2$.  The fields obtained along the duality orbit, seeded by the GMGHS solution are
\beq
E'_r =  \frac{Q}{r^2}e^{2\phi_\infty}\cos\beta \ , \qquad
D'_r = \frac{Q}{r(r-r_-)}\cos\beta \ ,
\label{ebho}
\eeq
\beq
B'_r = -\frac{Q}{r(r-r_-)}\sin\beta \ , \qquad
H'_r = -\frac{Q}{r^2}e^{2\phi_\infty}\sin\beta \ .
\label{bbho}
\eeq
Once again for $\beta=\pi/2$ we obtain a purely magnetic configuration in the dual model with
\beq
 f_{\pi/2}(\phi) = \frac{1}{f(\phi)} = e^{2\phi} \ , \qquad g(\phi)=0 \ .
\eeq
The magnetic induction and intensity are now non-trivial:
\beq
B'_r = -\frac{Q}{r(r-r_-)} \ , \qquad 
H'_r = f_{\pi/2}(\phi)B'_r = -\frac{Q}{r^2}e^{2\phi_\infty} \ ,
\eeq
whereas the electric intensity and induction are trivial
\beq
\mathbf{E}' = 0 = \mathbf{D}' \ .
\eeq
This magnetic dilatonic black hole configuration was first obtained by Garfinkle, Horowitz and Strominger in \cite{GarHoroStrom.1991}, wherein the electric configuration was actually obtained by this duality rotation. The electromagnetic duality transformation of the EMS model reduces, for this specific choice of $\beta$, to this simple example of S-duality in low energy string theory, amounting to the change $\phi \rightarrow -\phi$, which in this context is the dilaton field. 

We can just as easily find a dyon black hole for any other angle $\beta$, but in this case $g(\phi)$ becomes generically non-vanishing. As a concrete example take $\beta=\pi/4$. Then, the model along the duality orbit has $f_\beta=1/\cosh{2\phi}$, $g_\beta=-\tanh{2\phi}$ and its action is, explictly:
\beq
\mathcal{S}_{\frac{\pi}{4}}=\frac{1}{4\pi}\int d^4x \sqrt{-g} \left(\frac{R}{4}-\frac{1}{4\cosh{2\phi}}F'_{\mu\nu}F'^{\mu\nu}-\frac{\tanh{2\phi}}{4}F'_{\mu\nu}\tilde{F}'^{\mu\nu}-\frac{1}{2}\partial_\mu\phi \partial^\mu \phi\right) \ .
\label{action3}
\eeq
This model admits a dyonic black hole solution with the GMGHS geometry and scalar field, \eqref{lelGM} and (\ref{gmghss}), and the electromagnetic field \eqref{ebho}-\eqref{bbho} with $\beta=\pi/4$, which in covariant form reads:
\beq
{\bf F}=-\frac{Q'}{r^2}e^{2\phi_\infty}dt\wedge dr+Q'\sin\theta d\theta\wedge d\varphi \ \  \Leftrightarrow \  \  {\bf A}=- \frac{Q'}{r}e^{2\phi_\infty}dt-Q'\cos\theta  d\varphi \ ,
\eeq
where $Q'=Q/\sqrt{2}$. Comparing with \eqref{mm} one can confirm this describes an electric plus magnetic charge, a dyon. As far as we are aware, such closed form solution within model \eqref{action3} has not been discussed previously in the literature. Moreover, one can compute other exact, closed form solutions of this model, $e.g.$ rotating charged black holes.

\subsection{Other models with scalarised and axionic black holes}
Having understood the duality orbits, let us mention a set of other EMS models \eqref{action} wherein numerical black hole solutions have been constructed in the literature and whose duality orbits can be constructed. Examples include the following coupling functions:
\begin{enumerate}
	\item Exponential coupling: \ $f^\alpha_E(\phi) = e^{-\alpha\phi^2}$ \ , \ $g(\phi)=0$\ ;
	\item Power-law coupling: \  $f^\alpha_P(\phi) = 1 - \alpha\phi^2$ \ , \ $g(\phi)=0$\ ;
	\item Fractional coupling $f^\alpha_F(\phi) = \frac{1}{1+\alpha\phi^2}$ \ , \ $g(\phi)=0$\ ;
\item Higher power-law coupling $f^\alpha_{HP}(\phi) = 1 - \alpha\phi^4$ \ , \ $g(\phi)=0$\ ;
         \item Axionic coupling $f(\phi)=1$ \ , \ $g_A^\alpha=\alpha \phi$  \ .
     \item Axionic-type coupling $f(\phi)=1$ \ , \ $g_{AT}^\alpha=\alpha \phi^2$  \ .

\end{enumerate}
In all cases $\alpha$ is a coupling constant. Couplings 1-3 were discussed in \cite{CHetAl.SclChr.2018,PFetAl.SclChr.2019,DAetAl.EMSBH.2019} in the context of EMS models allowing spontaneous scalarisation of charged black holes (see also, $e.g.$~\cite{Myung:2018vug,Boskovic:2018lkj,Myung:2018jvi,Hod:2020ljo,Herdeiro:2019yjy,Brihaye:2019kvj,Myung:2019oua,
Konoplya:2019goy,Ikeda:2019okp,Zou:2019bpt,Ramazanoglu:2019jrr,Brihaye:2019gla,Blazquez-Salcedo:2019nwd,Zou:2019ays,Fernandes:2020gay}); all these coupling functions have the same behaviour for small values of $\alpha\phi^2$. Coupling 4 was discussed in~\cite{JBSetAl.EMSBH.2020}; it does not allow spontaneous scalarisations but it exhibits an interesting two-branch space of solutions with scalar hair, co-existing with the standard Reissner-Nordstr\"om black hole, in a trinity of non-uniqueness. Black holes with coupling 5 were first discussed in~\cite{Lee:1991jw} and revisited recently in~\cite{PFetAl.AxChr.2019}, wherein coupling 6 was also discussed, again in the context of spontaneous scalarisation of charged black holes. Various solutions for flat spacetime with coupling 5 were also found and discussed in \cite{Nikitin:2012} and \cite{Ouellet:2019}.


We shall not analyse the duality orbits for all these models in detail, but let us make some comments at the rotation point $\beta=\pi/2$. One can see the duality relates different couplings through the relations \eqref{fbetapi2}-\eqref{gbetapi2}. For instance, we get the following identities:
\beq
f^\alpha_{E\pi/2} = f^{-\alpha}_E \ ,
\label{des}
\eeq
\beq
f^{\alpha}_{P\pi/2} = f^{-\alpha}_F \ ,
\eeq
\beq
f^{\alpha}_{F\pi/2} = f^{-\alpha}_P \ .
\eeq
Thus, the exponential squared coupling enjoys a type of S-duality symmetry  analogous to that of the  dilatonic model of section~\ref{gmghs}, via~\eqref{des}, whereas  the power law and fractional couplings are along the same duality orbit, and can be mapped into  each other by also changing the sign of the coupling constant $\alpha$.
At $\beta=\pi/2$, moreover, the purely electric solutions of models 1-4, as before, become purely magnetic, with:
\beq
\mathbf{E}' = \mathbf{H} = 0 \ ,
\eeq
\beq \label{Bpi2}
\mathbf{B}' = - \mathbf{D} =  - f \mathbf{E} \ 
\eeq
These results are in agreement with the Bekenstein type identities found in \cite{DAetAl.EMSBH.2019}, where both $f_{,\phi\phi}$ and $\phi f_{,\phi}$ must have the opposite sign of $F_{\mu\nu}F^{\mu\nu}$ for solutions with a scalar profile to exist. For purely electric ($F^2<0$) or magnetic ($F^2>0$) solutions, these conditions imply a different sign for the coupling constant $\alpha$ for the couplings mentioned above.


\section{Conclusions}
Understanding the symmetries of any physical  theory is always of great importance. Electromagnetic duality is a symmetry  of the vacuum  Maxwell equations  which has led to important insights and  generalisations  in classical and quantum field theory, as well as in relativistic gravity. In this paper we have considered EMS models described by the action~\eqref{action} for which,  in general, electromagnetic duality rotations are not a symmetry of a specific model, but define an orbit in the space of EMS models, which encompasses all possible choices for the coupling functions  $f(\phi)$ and $g(\phi)$.  This orbit is a one parameter closed orbit. There can be fixed points of the duality action, which are self-dual theories. In our analysis, the only such point corresponds to Maxwell's theory, which,  in our  setup has $f=1$ and $g=0$. For this self-dual theory, the orbit shrinks down to a point. 

For self-dual theories, electromagnetic duality relates different solutions of the same theory. For non-self dual theories, electromagnetic duality relates different solutions of different theories. In either case, electromagnetic duality is a useful solution generating technique. In the case considered herein, the duality map generically relates models with different coupling functions and electromagnetic fields, leaving the scalar field and background geometry unchanged.

To illustrate how the duality orbits can be used as a solution generating technique we have considered some simple electrically charged solitonic and black hole solutions, obtaining the corresponding dyons along the duality orbit and, in particular, pure magnetic configurations that emerge at the particular rotation corresponding to $\beta=\pi/2$.
In these examples, the models had a vanishing coupling $g(\phi)$; but since $\tilde{F}_{\mu\nu}F^{\mu\nu}=0$ for these purely electric, spherically symmetric solutions, these are also solutions for any $g(\phi)$ coupling one may  choose. A different orbit of solutions  exists for each possible $g(\phi)$. We have also obtained a new dyonic black hole solution of the model~\eqref{action3}, which illustrates the usefulness of this technique.

As a direction of further research, it could  be instructive to use  this approach for a two field model,  inspired  by the  dilaton-axion  model  in \cite{GibRas1996}, but with general couplings instead of the typical dilaton-axion couplings.

\section*{Acknowledgement}
We would like to thank E. Radu for many useful discussions. J.O. is supported by the FCT grant PD/BD/128184/2016.
This  work  is  supported  by  the  Center  for  Research  and
Development  in  Mathematics  and  Applications  (CIDMA)  and  by
the  Centre  of  Mathematics  (CMAT)  through  the  Portuguese  Foundation  for  Science  and  Technology  (FCT  -  Funda\c c\~ao  para  a  Ci\^encia e a Tecnologia), references UIDB/04106/2020, UIDP/04106/2020,
UIDB/00013/2020 and UIDP/00013/2020.  We  acknowledge support from the projects PTDC/FIS-OUT/28407/2017 and  CERN/FIS-PAR/0027/2019. This work
has  further  been  supported  by  the  European  Union's  Horizon  2020
research  and  innovation  (RISE)  programme  H2020-MSCA-RISE-2017
Grant no. FunFiCO-777740. The authors would like to acknowledge
networking support by the COST Action CA16104.

\end{document}